# Reduction of valuation risk by Kalman filtering in business valuation models

**René Scheurwater**
Dutch Health Authority,
Burgemeester Geurtsweg 5,
6065 EE Montfort,
The Netherlands

## Abstract
A recursive free cash flow model (FCFF) is proposed to determine the corporate value of a company in an efficient market in which new market and company-specific information is modelled by additive white noise. The stochastic equations of the FCFF model are solved explicitly to obtain the average corporate value and valuation risk. It is pointed out that valuation risk can be reduced significantly by implementing a conventional two-step Kalman filter in the recursive FCFF model, thus improving its predictive power. Systematic errors of the Kalman filter, caused by intermediate changes in risk and hence in the weighted average cost of capital (WACC), are detected by measuring the residuals. By including an additional adjustment step in the conventional Kalman filtering algorithm, it is shown that systematic errors can be eliminated by recursively adjusting the WACC. The performance of the three-step adaptive Kalman filter is tested by Monte Carlo simulation which demonstrates the reliability and robustness against systematic errors. It is also proved that the conventional and adaptive Kalman filtering algorithms can be implemented into other valuation models such as the economic value added model (EVA™) and free cash flow to equity model (FCFE).

## 1. Introduction
Research during the last two decades of the 20$^{th}$ century pointed out that cash flow based measures better correlate with stock prices than traditional accounting earnings-based measures such as earnings per share (EPS) and Price-to-Earning (P/E) [15]. Since stock prices tend to reflect market expectation of the future returns, growth and risk of a company, investors and managers increasingly looked at cash flow measures as indicators of corporate value and shareholder value. This led to development of Value Based Management (VBM), a framework for measuring the value drivers of the company and translating these into operational measures in order to enhance shareholder value, thus establishing a link between strategic management and operational management [2, 4, 13].

Within the VBM framework, the corporate value of a company is calculated as the present value of future free cash flows to the firm (FCFF) discounted at the weighted average cost of capital (WACC). The free cash flow FCFF is the cash flow available to all the investors of capital in the firm (common stockholders, preferred stockholders, bondholders and other claimholders) after investment in working capital and fixed capital and payment of all operating expenses and taxes. The market value of equity or shareholder value of the company is readily obtained from the corporate value by subtracting the market value of debt (preferred stock, bonds and other claims). WACC is the opportunity cost for investing in the debt and equity of a company, determined by the capital-asset-pricing model (CAPM) as the weighted average of the risk-adjusted rate of return on debt and equity [14].

The fundamental principle of VBM is that companies create shareholder value by delivering products or services thereby achieving a return on invested capital which exceed the opportunity cost WACC. Since creation of shareholder value is the main objective for managers and investors, discounted cash flow models have been developed in the past decades for almost all industry sectors in order to simulate alternative corporate strategies in case of mergers, acquisitions, divestitures, start-ups, capital restructuring and business turn-around situations [5, 14]. Most of these discounted cash flow models are sector-specific generalizations of the Free Cash Flow to the Firm (FCFF) [4, 13], Free Cash Flow to Equity (FCFE) [5], Economic Value Added (EVA™) [2, 12, 16], Cash Flow Return On Investment (CFROI®) ‘[10].

Uncertainty in the modelling the future free cash flows due to new market or company-specific information, inevitably leads to uncertainty in the outcomes of discounted cash flow models i.e. valuation risk [3]. It will be proved in this paper that valuation risk can be reduced significantly by implementing a conventional Kalman filter into the discounted cash flow models thus improving reliability and predictive power of discounted cash flow models.

In conventional Kalman filters the information of a process model is combined with the outcomes of a measurement model and filtering is carried out by a two-step recursive algorithm [1, 7, 8]. In the first step (the “value prediction step”) the outcome at time t is predicted by the recursive process model by using the updated value at the previous time step t-1. In the second step (the “measurement update step”) the predicted value is updated by real-time measurement of the process at time t. The updated value is calculated as the weighted average of the predicted value and the measurement residual, where the weight factor is called the Kalman gain parameter. The main reason that Kalman filtering is a popular and powerful estimation method is that the update step only requires information about the previous time step rather than the entire previous history and also that it is straightforward to determine the optimal Kalman gain parameter at which the variance of the updated error estimate is minimized. This explains the numerous applications of Kalman filters in real-time data processing and navigation (such as robotics, global positioning systems, missile trajectory optimization) [9].

The accuracy of a conventional Kalman filter strongly depends on the parameters and noise statistics of the process model and measurement model. In most applications model parameters and noise statistics are determined a priori and remain unchanged during the filtering process. Since the performance of the conventional Kalman filter is very sensitive to changes in the process and measurement models, it is only allowed to keep the model parameters and noise statistics constant during the filtering process as long as they match the actual models. In case the a priori determined model parameters and noise levels are inadequate to represent the actual process model and measurement model, conventional Kalman filters may lead to unreliable results and systematic errors. A common solution to avoid systematic errors is to assign relative large noise levels, but this will reduce the performance of the conventional Kalman filter. An alternative approach to eliminate systematic errors is to use exploit an adaptive Kalman filter which adjusts the model parameters or noise statistics during the filtering process [6, 11].

The outline of this paper is as follows. In Section 2 a recursive FCFF model is given for valuation of a company in an efficient market in which new market and company-specific information is being modelled by additive white noise. Recursive equations are solved explicitly in order to derive expressions for the average corporate value and valuation risk benchmark. The expression of the valuation risk benchmark, defined as the present value of the variances of future value fluctuations, will be adopted as basic risk measure in following sections of this paper. In Section 3 a conventional Kalman filter is constructed by adopting the FCFF model as process model which is combined with a measurement model based on the available market information. Recursive equations are derived for the variances of the error estimates of value prediction step and measurement update step. The optimal Kalman gain parameter is determined at which the variance of the updated error estimate is minimized. In Section 4 Monte Carlo simulations show that valuation risk with Kalman filtering can be significantly lower than the valuation risk benchmark calculated in Section 2 without Kalman filtering. In Section 5 model parameters are determined at which valuation risk with Kalman filtering is being minimized. In Section 6 systematic errors in Kalman filtering due to intermediate changes in risk and hence in weighted average cost of capital (WACC) are discussed and it is pointed out that systematic errors can be detected during the filtering process by measuring the residuals. In Section 7 an adaptive Kalman filtering algorithm is proposed which not only eliminates these systematic errors but also adjusts the predicted cost of capital to approximate the actual cost of capital. In Section 8 a Monte Carlo simulation is carried out which demonstrates the reliability and robustness of the adaptive Kalman filtering algorithm to changes in market conditions. In the Appendix it is pointed out that the Kalman filtering algorithm also can be implemented into other valuation models such as the EVA™ and the FCFE valuation models.

## 2. Stochastic FCFF model for a company in an efficient market

Given the corporate value $V_{t-1}$ and free cash flow to the firm $FCFF_{t-1}$ at time t-1, the corporate value $V_t$ at time t can be determined by the following recursive formula

$$V_t = (1 + R_t).V_{t-1} - FCFF_{t-1} + \sigma_t.\varepsilon_t \qquad (2.1)$$

where $R_t$ is the weighted average cost of capital (WACC) at time t. According to the Efficient Market Hypothesis (EMH) stock prices in an efficient and competitive capital market fully reflect available market and company-specific information and new information is incorporated instantaneously in stock prices and corporate value $V_t$ [14]. The effect of new market or company-specific information on the corporate value $V_t$ is modelled by additive white noise $\varepsilon_t$ in Eq. (2.1). Without loss of generality the stochastic fluctuations $\varepsilon_t$ can be normalized such that

$$Avg(\varepsilon_t) = 0, \quad Var(\varepsilon_t) = 1, \; Covar(\varepsilon_{t+n}, \varepsilon_t) = \delta_{n,0} \qquad (2.2)$$

With this normalization $\sigma_t$ can be identified with the standard deviation of the random fluctuations in the company value at time t. It is assumed that the free cash flow $FCFF_{t-1}$ realized over the time period between time step t-1 and time step t is entirely determined by the market specific and company-specific information available at time t-1 and not yet affected by new information arriving at time t. With this assumption the FCFF model Eq. (2.1) reduces to a univariate model for $V_t$ with model parameters $FCFF_{t-1}$, $R_t$ and $\sigma_t$. The main reason to make this assumption is to keep the analytical steps for implementing the Kalman filtering algorithm as simple as possible. Obviously, the valuation model Eq. (2.1) can be generalized by including a recursive relation for the stochastic variable $FCFF_t$ in Eq. (2.1) thus leading to a coupled bivariate FCFF model for $F_t$ and $V_t$. Despite the implementation steps of the Kalman filtering algorithm in a bivariate FCFF model are the same, the analysis will be more complex and therefore will be discussed in a forthcoming paper.

**Recursion equation for the average corporate value**

In order to solve the univariate FCFF model Eq. (2.1) and derive expressions for the statistical moments of $V_t$, it turns out to be convenient to decompose $V_t$ into an average part $Avg(V_t)$ and a fluctuating part $\delta V_t$,

$$V_t = Avg(V_t) + \delta V_t \qquad (2.3)$$

Inserting the decomposition Eq. (2.3) into Eq. (2.1) and recalling that $R_t$ and $FCFF_t$ are deterministic functions of t, the recursion equation for $Avg(V_t)$ is given by

$$Avg(V_t) = (1 + R_t).Avg(V_{t-1}) - FCFF_{t-1} \qquad (2.4)$$

By solving Eq. (2.4) recursively, the average corporate value $Avg(V_t)$ can be expressed in terms of $FCFF_t$ to obtain

$$Avg(V_t) = \frac{FCFF_t}{(1 + R_{t+1})} + \frac{FCFF_{t+1}}{(1 + R_{t+1}).(1 + R_{t+2})} + \cdots \qquad (2.5)$$

Therefore, the average corporate value $Avg(V_t)$ equals the present value of all future free cash flows $FCFF_t$, $FCFF_{t+1}$, …. thus showing that the recursive model Eq. (2.1) is equivalent to the FCFF model

**Recursion equation for the value fluctuations**

Inserting Eq. (2.3) into Eq. (2.1) and subtracting Eq. (2.4), the recursive equation for the fluctuating part $\delta V_t$ is given by

$$\delta V_t = (1 + R_t).\delta V_{t-1} + \sigma_t.\varepsilon_t \qquad (2.6)$$

By recursive application it follows from Eq. (2.6) the value fluctuation $\delta V_t$ can be expressed in terms the future market fluctuations $\varepsilon_{t+1}, \varepsilon_{t+2}, \ldots$

$$\delta V_t = -\frac{\sigma_{t+1}.\varepsilon_{t+1}}{(1+R_{t+1})} - \frac{\sigma_{t+2}.\varepsilon_{t+2}}{(1+R_{t+1}).(1+R_{t+2})} - \ \dots$$

(2.7)

The variance $Var(\delta V_t)$ can be obtained from Eq. (2.7) by exploiting the white noise conditions Eq. (2.2)

$$Var(\delta V_t) = \frac{{\sigma_{t+1}}^2}{(1+R_{t+1})^2} + \frac{{\sigma_{t+2}}^2}{(1+R_{t+1})^2.(1+R_{t+2})^2} + \cdots$$

(2.8)

hence the variance $Var(\delta V_t)$ equals the present value of the variances ${\sigma_t}^2$ of future value fluctuations due to new market or company-specific information. Defining valuation risk at time t as the variance $Var(\delta V_t)$ given by Eq. (2.8), then valuation risk for the univariate valuation model Eq. (2.1) is independent of $FCFF_t$ and entirely determined by the value fluctuations due to new market and information at future times t+1, t+2, …..

**<u>Example – Statistical moments for a 1-period FCFF model</u>**

As an example, consider a mature company with constant free cash flow $FCFF_t = F$ in a stable market with stationary noise statistics $\sigma_t = \sigma$. Also assuming that the cost of capital is constant $R_t = R$, then since all parameters are constant then the statistical moments will also be constant (i.e. constant average corporate value and constant valuation risk) and the firm is said to be in steady state (or going concern). For this so-called 1-period FCFF model the expression Eq. (2.5) for the average corporate value reduces to the perpetuity

$$Avg(\mathrm{V_t}) = \frac{F}{R}$$

(2.9)

whereas the expression Eq. (2.8) for the valuation risk simplifies to

$$Var(\delta \mathrm{V_t}) = \frac{\sigma^2}{R^2 + 2R}$$

(2.10)

for all $t \geq 0$. Here the sum formula has been exploited

$$\sum_{k=1}^{H} x^k = \frac{x - x^{H+1}}{1-x}$$

(2.11)

with $x = 1/(1+R)$. It readily follows from Eq. (2.7) and the white noise conditions Eq. (2.2) that the covariance, which is a measure for the interdependence of value fluctuations at different times, is given by

$$Covar(\delta V_t, \delta V_{t+n}) = \frac{1}{(1+R)^n} . \frac{\sigma^2}{R^2 + 2R}$$

(2.12)

Therefore, the correlation function $Corr(\delta V_{t+n}, \delta V_t)$ for value fluctuations $\delta V_t$, defined as

$$Corr(\delta V_{t+n}, \delta V_t) = \frac{Covar(\delta V_{t+n}, \delta V_t)}{Var(\delta V_t)} = \frac{1}{(1+R)^n}$$

(2.13)

only depends on the time lag $n$ and not on $t$, which is a consequence of the stationarity assumption of the random noise. In the next sections the valuation risk $Var(\delta V_t)$ given by Eq. (2.10) will be adopted as benchmark measure in order to compare the valuation risk of the FCFF model with Kalman filtering.

**Example – Statistical moments for a 2-period FCFF model**

Consider a 2-period FCFF model for a company with parameters $F_I, \sigma_I$ in period $I$ $(0 \leq t \leq H)$ and parameters $F_{II}, \sigma_{II}$ in period $II$ $(t > H)$ where $F_I > F_{II}, \sigma_I > \sigma_{II}$. Let the weighted average cost of capital $R_t = R$ be constant for all $t \geq 0$ then with boundary conditions at the forecast horizon $t = H$,

$$Avg(V_H) = \frac{F_{II}}{R}, \qquad Var(\delta V_H) = \frac{{\sigma_{II}}^2}{R^2 + 2R}$$

(2.14)

the solution Eqs. (2.5) and (2.8) for $0 \leq t \leq H$ are given by

$$Avg(V_t) = \frac{F_I}{R} + (1+R)^{t-H}.\frac{(F_{II} - F_I)}{R}$$

(2.15)

$$Var(\delta V_t) = \frac{{\sigma_I}^2}{R^2 + 2R} + (1+R)^{2(t-H)}.\frac{({\sigma_{II}}^2 - {\sigma_I}^2)}{R^2 + 2R}$$

where sum formula Eq. (2.11) has been used.

In Fig. 1 ten different Monte Carlo realizations are given of Eq. (2.1) for a 2-period FCFF model with forecast horizon $H = 20$ and parameters $F_I = 10{,}0, R_I = 0{,}1, \sigma_I = 1{,}0$ in period I $(0 \leq t \leq 20)$ and parameters $F_{II} = 7{,}0$, $R_{II} = 0{,}1, \sigma_{II} = 0{,}7$ in period II $(t > 20)$. Clearly, the corporate value $V_t$ gradually decreases between t=0 and t=20, due to the fact that closer to the horizon $H = 20$, the number of years with large values $F_I$ decreases. From the boundary conditions Eq. (2.14) it follows that the average $Avg(V_H) = F_{II}/\ R_{II} =$ 70,0 and variance is $Var(\delta V_H) = {\sigma_{II}}^2/({R_{II}}^2 + 2R_{II})\ = 2{,}333$ for t $> 20$. The upper and lower 95%-significance levels are given by $Avg(V_H) \pm 1{,}96.Stdev(\delta V_H) = Avg(V_H)) \pm$ 2,994.

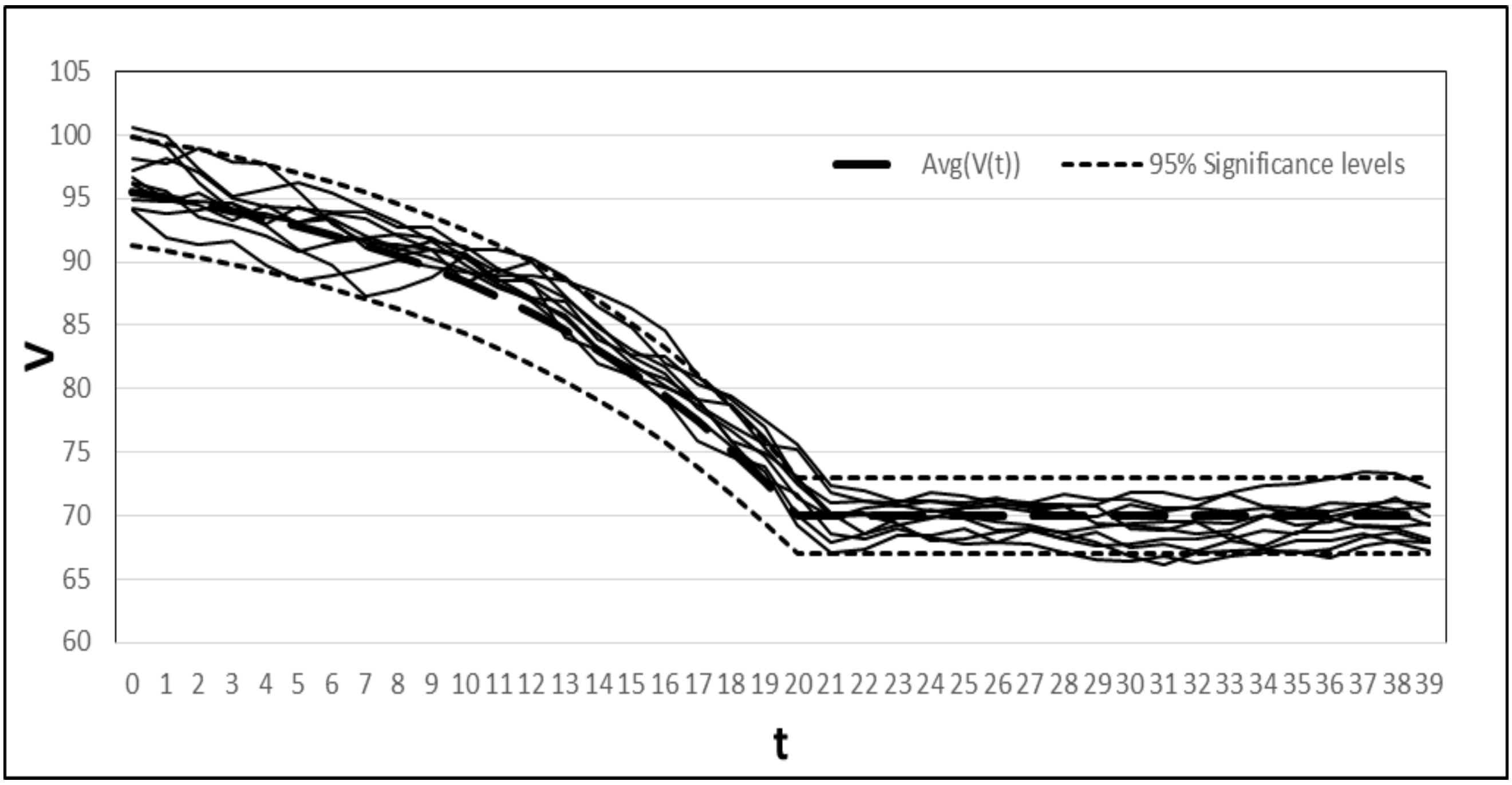

Fig 1. Monte Carlo simulations of Eq. (2.1) for $V_t$ and the upper and lower 95%-significance levels.

## 3. Implementation of a conventional Kalman filter in the FCFF model

In this section it will be pointed out that valuation risk $Var(\delta V_0)$, can be significantly reduced by implementing a Kalman filtering algorithm in the valuation model Eq. (2.1). The recursive Kalman filtering algorithm is carried out in two steps: 1) the "value prediction step" in which the company value $V_t$ at time t is predicted by the valuation model Eq. (2.1) and 2) the "measurement update step" in which the predicted company value at time t is updated by measuring the available corporate and market information at time t.

**1. Value prediction step**

Given the updated company value $V_{t-1|t-1}$ at time t-1, then the predicted company value $V_{t|t-1}$ at time t is determined as follows

$$V_{t|t-1} = (1 + R_t).V_{t-1|t-1} - FCFF_{t-1} \qquad (3.1)$$

**2. Measurement update step**

After carrying out the prediction step at time t, the actual but unknown corporate value $V_t$ is estimated by measuring the market value (such as stock price, multiples etc.). For the linear Kalman filter the outcome of the measurement $W_t$ is linearly related to the actual company value $V_t$ by some scaling parameter $h$,

$$W_t = h.V_t + \lambda_t.\omega_t \qquad (3.2)$$

The measured value $W_t$ is allowed to be noisy with measurement error $\omega_t$ which is assumed to be white noise. Without loss of generality the parameter $\lambda_t$ can be normalized such that

$$Avg(\omega_t) = 0, \quad Var(\omega_t) = 1, \; Covar(\omega_{t+n}, \omega_t) = \delta_{n,0} \qquad (3.3)$$

and the parameter $\lambda_t$ can be identified with the standard deviation of the measurement noise. It is further assumed that the measurement noise $\omega_t$ is not correlated with the process noise $\varepsilon_t$,

$$Covar(\varepsilon_{t+m}, \omega_{t+n}) = 0 \text{ for all m,n} \quad (3.4)$$

In a linear Kalman filter algorithm, the updated company value $V_{t|t}$ is given by the weighted average of the predicted value $V_{t|t-1}$ and the measurement residual $Res_t = W_t - h.V_{t|t-1}$,

$$V_{t|t} = V_{t|t-1} + k_t.\left(W_t - h.V_{t|t-1}\right) \quad (3.5)$$

where the weight $k_t$ is called the Kalman gain parameter. In section 4 it will be shown that $0 \leq h.k_t \leq 1$. Clearly, the larger the $k_t$ value, the more weight is put on the outcome of the measurement and less on the prediction by the valuation model. Indeed, in case $h.k_t \approx 1$, it follows from Eq. (3.5) that the Kalman filter places all weight on the measurement $W_t$ and consequently $V_{t|t} \approx W_t/h$, whereas in case $h.k_t \approx 0$ the Kalman filter puts all weight on the prediction $V_{t|t-1}$ and consequently $V_{t|t} \approx V_{t|t-1}$.

**Recursion equations for the error estimates**

From Eq. (2.1) for the actual company value $V_t$ and the predicted value given by Eq. (3.1), the predicted error $E_{t|t-1} \equiv V_t - V_{t|t-1}$ is

$$E_{t|t-1} = (1 + R_t).E_{t-1|t-1} + \sigma_t.\varepsilon_t \quad (3.6)$$

Inserting Eq. (3.2) into Eq. (3.5) then the updated error $E_{t|t} \equiv V_t - V_{t|t}$

$$E_{t|t} = (1 - h.k_t).E_{t|t-1} - k_t.\lambda_t.\omega_t \quad (3.7)$$

Note that Eqs. (3.6) and (3.7) constitute a coupled system of recursion equations for the error estimates $E_{t|t-1}$ and $E_{t|t}$. Inserting Eq. (3.6) into Eq. (3.7) the predicted error $E_{t|t-1}$ can be eliminated so as to obtain a single recursion equation for the updated error $E_{t|t}$.

$$E_{t|t} = (1 - h.k_t).(1 + R_t).E_{t-1|t-1} + (1 - h.k_t).\sigma_t.\varepsilon_t - k_t.\lambda_t.\omega_t \quad (3.8)$$

By starting the recursion algorithm at time t=0 with initial condition $E_{0|0} = 0$, it readily follows from Eq. (3.8) that $E_{t|t}$ is a linear function of the realizations $\varepsilon_1, \varepsilon_2, \dots, \varepsilon_t$ and $\omega_1, \omega_2, \dots, \omega_t$ hence by Eq. (3.6) $E_{t|t-1}$ is a linear function of the realizations $\varepsilon_1, \varepsilon_2, \dots, \varepsilon_t$ and $\omega_1, \omega_2, \dots, \omega_{t-1}$. Recalling the white noise conditions Eqs. (2.2), (3.3) and (3.4)

$$Covar\left(E_{t-1|t-1}, \varepsilon_t\right) = 0,\ Covar\left(E_{t-1|t-1}, \omega_t\right) = 0,\ Covar\left(E_{t|t-1}, \omega_t\right) = 0 \quad (3.9)$$

**Recursion equations for the variance of the error estimates**

By taking the square of Eq. (3.6) and applying the white noise conditions Eqs. (3.9), a recursion equation can be derived for the predicted error variance $Var\left(E_{t|t-1}\right)$,

$$Var(E_{t|t-1}) = (1 + R_t)^2.Var(E_{t-1|t-1}) + {\sigma_t}^2 \quad (3.10)$$

Similarly, by squaring Eq. (3.7) and applying the white noise conditions Eqs. (3.9) a recursion relation is obtained for the updated error variance $Var(E_{t|t})$

$$Var(E_{t|t}) = (1 - h.k_t)^2.Var(E_{t|t-1}) + (k_t.\lambda_t)^2 \tag{3.11}$$

Inserting Eq. (3.10) into Eq. (3.11) the predicted error variance $Var(E_{t|t-1})$ can be eliminated so as to obtain a recursion equation entirely expressed in terms of the updated error covariance $Var(E_{t|t})$

$$Var(E_{t|t}) = a_t{}^2.Var(E_{t-1|t-1}) + b_t \tag{3.12}$$

where the coefficients $a_t, b_t$ are defined as

$$a_t = (1 - h.k_t).(1 + R_t)$$

$$b_t = (1 - h.k_t)^2.\sigma_t{}^2 + (k_t.\lambda_t)^2 \tag{3.13}$$

**<u>Calculation of the optimal Kalman gain parameter</u>**

In previous sections the Kalman gain $k_t$ has been treated as an arbitrary parameter. It will now be pointed out that the Kalman gain can be determined so as to minimize the variance of the updated error estimate $Var(E_{t|t})$ i.e. valuation risk. Minimization of $Var(E_{t|t})$ requires that the following two conditions are obeyed

$$\partial Var(E_{t|t})/\partial k_t = 0$$

$$\partial^2 Var(E_{t|t})/\partial k_t{}^2 > 0 \tag{3.14}$$

By taking the first derivative of Eq. (3.11) with respect to $k_t$, the first condition of Eq. (3.14) is satisfied for $k = k_{0,t}$, where

$$h.k_{0,t} = \frac{\mathrm{h}^2.Var(E_{t|t-1})}{\mathrm{h}^2.Var(E_{t|t-1}) + \lambda_t{}^2} \tag{3.15}$$

Since $Var(E_{t|t-1}) > 0$ it follows that $0 \leq h.k_{0,t} \leq 1$. The second condition of Eq. (3.14) is trivially satisfied since by taking the second derivative of Eq. (3.11)

$$\frac{\partial^2 Var(E_{t|t})}{\partial k_t{}^2} = 2\lambda_t{}^2 + 2\mathrm{h}^2.Var(E_{t|t-1}) > 0 \tag{3.16}$$

This proves that the variance of the updated error variance $Var(E_{t|t})$ is indeed minimized at the Kalman gain $k = k_{0,t}$ given by Eq. (3.15), which is therefore called the optimal Kalman gain. It turns out to be convenient to introduce the $Q_t$ ratio

$$Var(E_{t|t-1}) = (\lambda_t{}^2/\mathrm{h}^2).Q_t \tag{3.17}$$

Expressing the optimal Kalman gain Eq. (3.15) in terms of the $Q_t$ ratio

$$h.k_{0,t} = \frac{Q_t}{1+Q_t} \tag{3.18}$$

By inserting Eqs. (3.17) and (3.18) into Eq. (3.11) the minimum value for $Var(E_{t|t})$ is given by

$$Var(E_{t|t}) = ({\lambda_t}^2/\mathrm{h}^2).\frac{Q_t}{1+Q_t} = (1-h.k_{0,t}).Var(E_{t|t-1}) \tag{3.19}$$

The $Q_t$ ratio defined by Eq. (3.17) can be interpreted as a relative measure of the accuracy of the prediction process compared with the measurement process. This can be justified by considering two extreme cases. First consider the extreme case $Q_t \ll 1$ , then according to Eq. (3.17) the predicted variance $Var(E_{t|t-1})$ is much smaller than the measurement noise ${\lambda_t}^2/\mathrm{h}^2$ hence the measurement process is relatively inaccurate and the predicted company value will get more weight than the measurement in applying the Kalman algorithm. Consequently, from Eqs. (3.18) and (3.19) it then follows that $h.k_{0,\mathrm{t}} \ll 1$ and $Var(E_{t|t}) \approx Var(E_{t|t-1})$ for $Q_t \ll 1$. Next consider the extreme case $Q_t \gg 1$ , then from Eq. (3.17) it follows that predicted variance $Var(E_{t|t-1})$ is much larger than the measurement noise ${\lambda_t}^2/\mathrm{h}^2$ hence the measurement process is relatively accurate and in the Kalman filtering algorithm the measurement gets more weight than the predicted company value. This also follows from Eqs. (3.18) and (3.19) since $h.k_{0,\mathrm{t}} \approx 1$ and $Var(E_{t|t}) \ll Var(E_{t|t-1})$ for $Q_t \gg 1$.

## 4. Monte Carlo simulation of the FCFF model with conventional Kalman filtering

In order to show that the updated variance $Var(E_{t|t})$ with Kalman filtering can be significantly smaller than the valuation risk $Var(\delta V_t)$ without Kalman filtering, consider a 2-period FCFF model with forecast horizon H and parameters $F_I, \sigma_I$ in period I and parameters $F_{II}, \sigma_{II}$ in period II. It is furthermore assumed that $k_I = k_{II} = k,\ R_I = R_{II} = R,\ \lambda_I = \lambda_{II} = \lambda$. It can be verified that by exploiting the sum formula Eq. (2.11) the theoretical solution $Var(E_{t|t})$ of Eq. (3.12) is given by

$$Var(E_{t|t}) = \frac{b_I}{1-a^2} + a^{2t}.\left(Var(E_{0|0}) - \frac{b_I}{1-a^2}\right) \qquad 0 \le t \le H$$

$$Var(E_{t|t}) = \frac{b_{II}}{1-a^2} + a^{2(t-H)}.\frac{b_I - b_{II}}{1-a^2} + a^{2t}.\left(Var(E_{0|0}) - \frac{b_I}{1-a^2}\right) \qquad t \ge H \tag{4.1}$$

where $a = a_I = a_{II} = (1-h.k).(1+R)$ and $b_{I,II} = (1-h.k)^2.{\sigma_{I,II}}^2 + (k.\lambda)^2$. In Fig. 2 the outcomes are illustrated of a Monte Carlo simulation with 1000 different realizations of the 2-period FCFF model with and without Kalman filtering. It is assumed that the FCFF model has forecast horizon $H = 20$, cost of capital $R = 0{,}1$ for all $t \ge 0$ and valuation model parameters $F_I = 10{,}0,\ \sigma_I = 1{,}0$ in period I and $F_{II} = 7{,}0,\ \sigma_{II} = 0{,}7$ in period II. The measurement model parameters are $\lambda = 0{,}5,\ h = 1$ for all $t \geq 0$. Kalman filtering is carried out at a fixed value for the Kalman gain $k = 0{,}5$. In Fig. 2 also the theoretical values for $Var(\delta V_t)$ given by Eq. (2.15) and $Var(E_{t|t})$ given by Eq. (4.1) are given. The outcomes of the Monte Carlo simulation show that with the selected model parameters valuation risk $Var(E_{t|t})$ with Kalman filtering as given by Eq. (3.12) can be significantly smaller than the valuation risk $Var(\delta V_t)$ without Kalman filtering as given by Eq. (2.15).

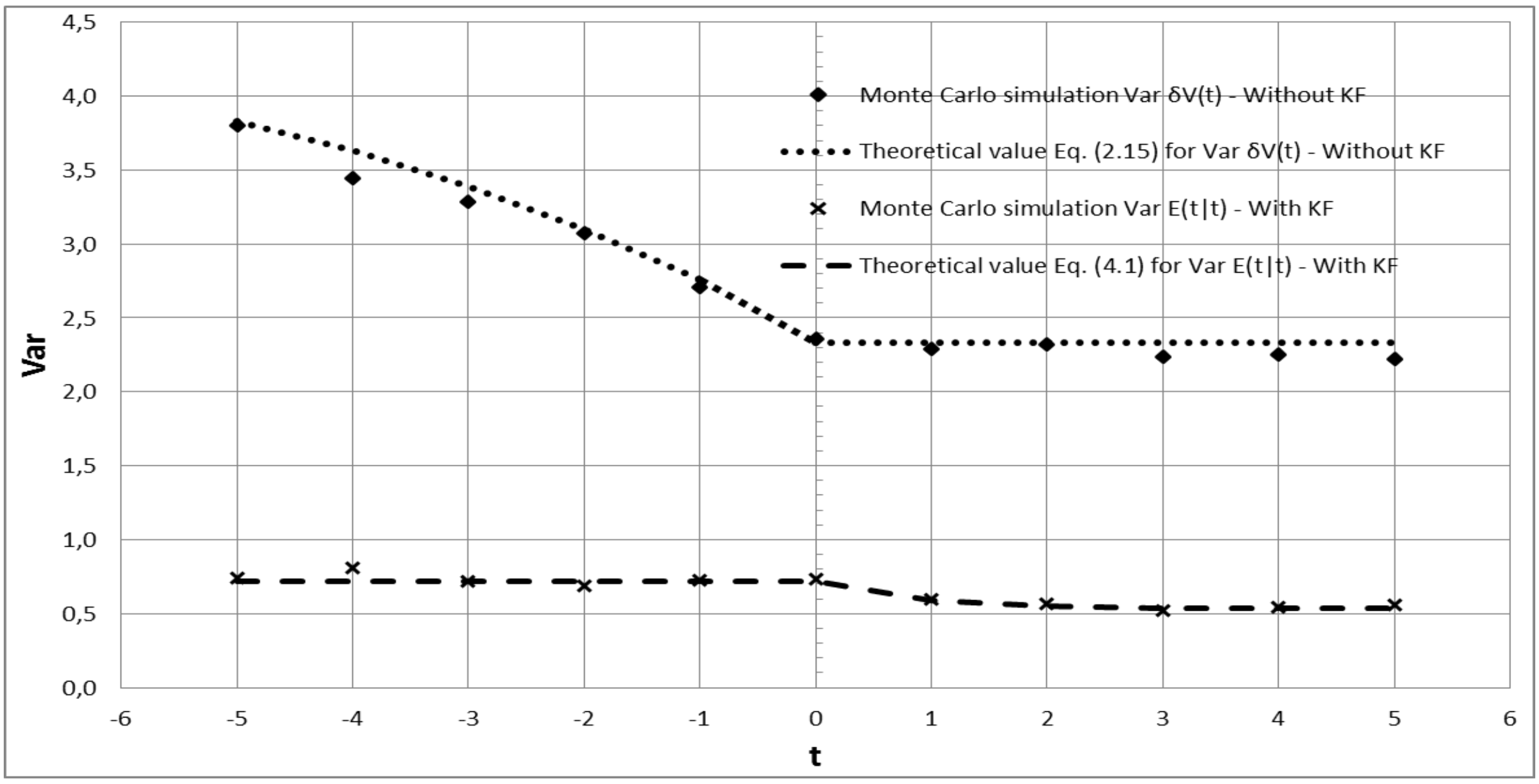

Fig 2. Monte Carlo simulation of the variance without Kalman filtering Eq. (2.15) and with Kalman filtering Eq. (4.1)

## 5. Minimization of valuation risk in the FCFF model with conventional Kalman filtering

In order to determine the model parameters for which valuation risk Eq. (3.12) is smaller than the benchmark measure Eq. (2.12), consider a company in a state of going concern with constant coefficients $k_t = k, R_t = R,\ \lambda_t = \lambda, \sigma_t = \sigma$. Recalling Eqs. (4.1) then in the limit $t \to \infty$ the updated error variance $Var(E_{t|t})$ asymptotically converges to the limit value $Var(E_{\infty|\infty})$ given by

$$Var(E_{\infty|\infty}) = \frac{b}{1-a^2} = \frac{(1-h.k)^2.\sigma^2 + (h.k)^2.(\lambda/h)^2}{1-(1-h.k)^2.(1+R)^2} \tag{5.1}$$

which by Eq. (3.18) can be rewritten as

$$Var(E_{\infty|\infty}) = \frac{\sigma^2 + (\lambda/h)^2.Q^2}{(1+Q)^2 - (1+R)^2} \tag{5.2}$$

An alternative expression for the variance $Var(E_{\infty|\infty})$ of the updated error estimate at the optimal Kalman gain readily follows from Eq. (3.19),

$$Var(E_{\infty|\infty}) = (\lambda/h)^2 . \frac{Q}{1+Q}$$

(5.3)

Since the RHS terms of Eqs. (5.2) and (5.3) must be equal, it is straightforward to verify that the Q factor obeys the quadratic equation

$$x^2 . Q^2 - \Pi . Q - 1 = 0 \qquad (5.4)$$

where $\Pi \equiv 1 + x^2 . (2\mathrm{R} + \mathrm{R}^2)$ and $x = \lambda/h\sigma$ is the dimensionless ratio of the measurement noise and process noise. The solution $Q$ of the quadratic equation is Eq. (5.4) is given by

$$Q = \frac{\Pi + \sqrt{\Pi^2 + 4x^2}}{2x^2}$$

(5.5)

The second solution of the quadratic Eq. (5.4) can be neglected since $Q \geq 0$. Note that the solution $Q$ given by Eqs. (5.5) is a function of the cost of capital $R$ and the ratio $x$ only.

Recalling that convergence of the $Var(E_{t|t})$ in Eq. (3.12) requires that $a = (1 - h.k).(1 + R) < 1$. At the optimal Kalman gain Eq. (3.18) the condition for convergence becomes $a = (1 + R)/(1 + \ Q) < 1$ hence $Q > R$. In Appendix A it is shown that at the optimal Kalman gain $Q > R$ hence $a < 1$ for all values $x = \lambda/h\sigma \geq 0$. In Fig. 3 the ratio $Q$ given by Eq. (5.4) is illustrated as a function of $x$ and different arbitrary but fixed values of $R$.

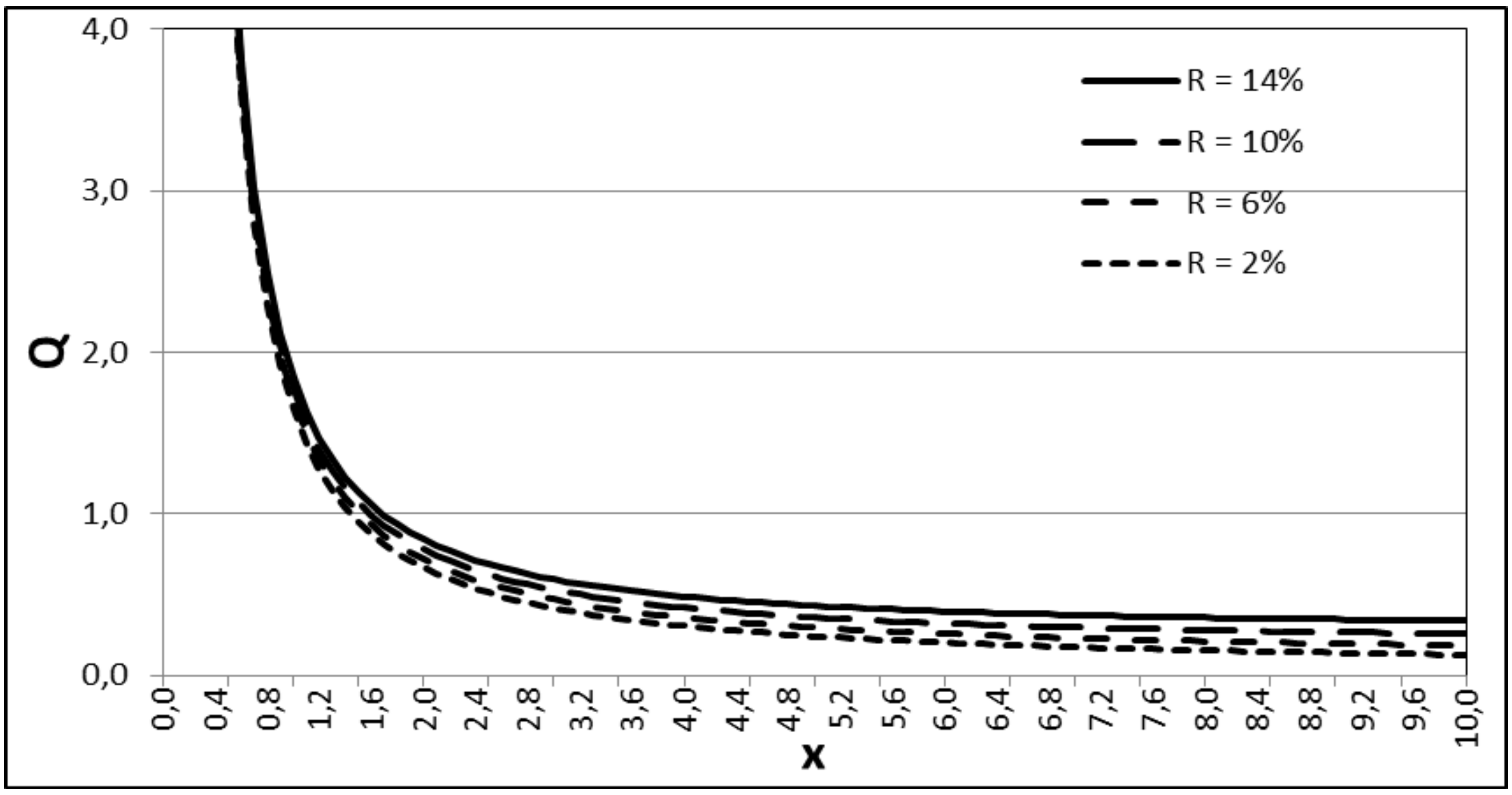

Fig 3. Ratio $Q$ as a function of cost of capital $R$ and parameter $x = \lambda/h\sigma$.

Inserting the solution Eq. (5.5) into Eq. (3.18) the optimal Kalman gain $k = k_0$ is given by

$$h.k_0 = \frac{Q}{1+Q} = \frac{2}{2 - \Pi + \sqrt{\Pi^2 + 4x^2}}$$

(5.6)

Note that the optimal Kalman gain $h.k_0$ given by Eq. (5.6) is a function of the cost of capital $R$ and the parameter $x = \lambda/h\sigma$. In Fig. 4 the optimal Kalman gain Eq. (5.6) is illustrated as a function of $x$ for some fixed values of $R$.

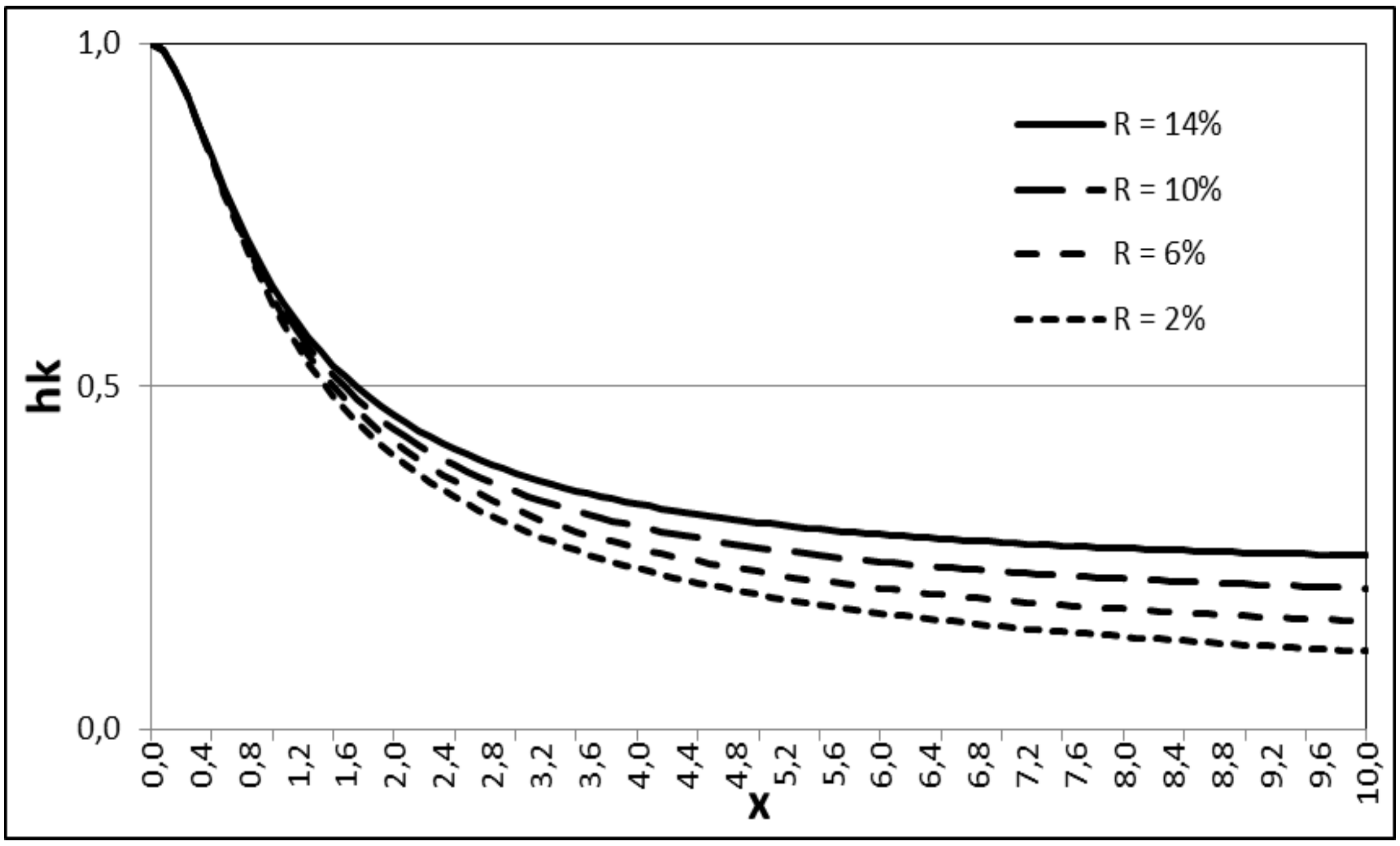

Fig 4. Optimal Kalman gain $h.k_0$ as a function of cost of capital $R$ and parameter $x = \lambda/h\sigma$

Substituting the solution Eq. (5.5) back into Eq. (5.3) the limit value $Var(E_{\infty|\infty})$ at the optimal Kalman gain is

$$Var(E_{\infty|\infty}) = \sigma^2 . \frac{2.x^2}{2 - \Pi + \sqrt{\Pi^2 + 4x^2}}$$

(5.7)

In order to compare the valuation risk Eq. (5.7) with the benchmark measure Eq. (2.10) without Kalman filtering, define the risk ratio $\mathcal{R}$,

$$\mathcal{R} = \frac{Var(E_{\infty|\infty})}{Var(\delta V_0)}$$

(5.8)

which upon inserting Eqs. (2.10) and (5.8) gives

$$\mathcal{R} = x^2.(R^2 + 2R).\frac{Q}{1+Q} = \frac{2.x^2.(R^2 + 2R)}{2 - \Pi + \sqrt{\Pi^2 + 4x^2}}$$

(5.9)

It is clear that the risk ratio $\mathcal{R}$ is minimized at $Q$ given by Eq. (5.5). This can be verified by differentiating Eq. (5.2) with respect to $Q$ and putting the derivative $\partial Var(E_{\infty|\infty})/\partial Q$ equal to zero, thus retaining the quadratic equation Eq. (5.4) with solution (5.5).

In Fig. 5 the risk ratio $\mathcal{R}$ is given as a function of $x = \lambda/h\sigma$ for several parameters $R$. Note that $\mathcal{R} < 1$ over a relatively large range of $x$ values, which proves the effectiveness of the Kalman filtering in reducing the valuation risk, even for values of $x > 1$ for which the measurement is less accurate than the prediction.

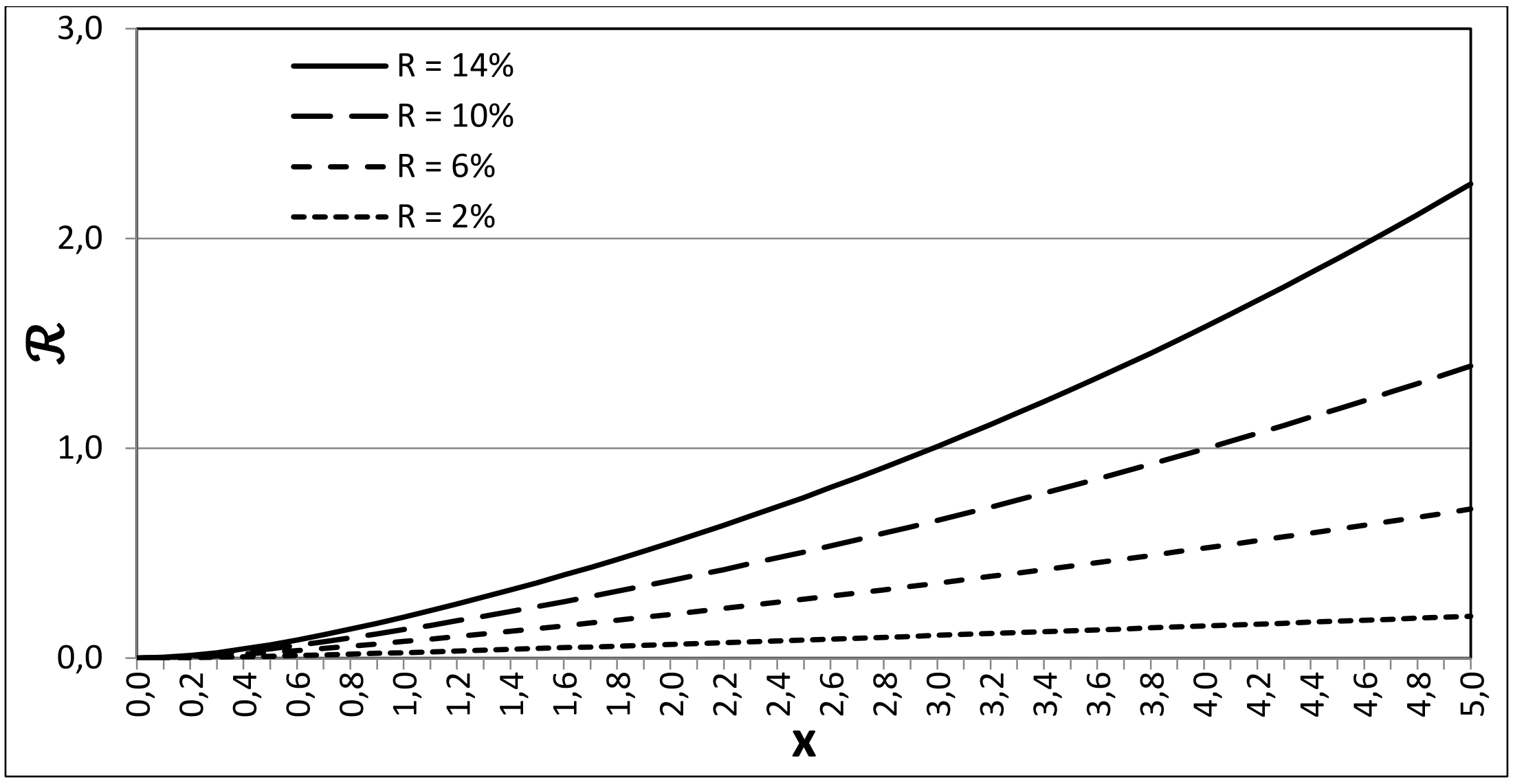

Fig 5. Risk ratio $\mathcal{R}$ as a function of cost of capital $R$ and parameter $x = \lambda/h\sigma$.

In Table 1 the main formulas are summarized for the recursive FCFF model with two-step Kalman filtering algorithm.

| |
|---|
| Step 1. Value prediction step<br><br>(3.1) $V_{t\|t-1} = (1 + R_t).V_{t-1\|t-1} - FCFF_{t-1}$<br><br>(3.10) $Var(E_{t\|t-1}) = (1 + R_t)^2.Var(E_{t-1\|t-1}) + {\sigma_t}^2$ |
| Step 2. Measurement update step<br><br>(3.5) $V_{t\|t} = V_{t\|t-1} + k_{0,t}.(W_t - h.V_{t\|t-1})$<br><br>(3.19) $Var(E_{t\|t}) = (1 - h.k_{0,t}).Var(E_{t\|t-1})$<br><br>(3.15) $h.k_{0,t} = \frac{\mathrm{h}^2.Var(E_{t\|t-1})}{\mathrm{h}^2.Var(E_{t\|t-1}) + {\lambda_t}^2}$ |

Table 1. Summary of the recursive formulas for the FCFF model with two-step Kalman filtering.

## 6. Systematic errors of the FCFF model with conventional Kalman filtering

The standard (also called conventional) two-step Kalman filter algorithm discussed in previous sections, is the optimal estimation method as long as the following conditions are satisfied

1) the valuation model is linear and the average value $V_{t|t-1}$ perfectly matches the actual average value $V_t$;
2) the value fluctuations $\varepsilon_t$ and measurement noise $\omega_t$ are uncorrelated white noise with zero average $Avg(\varepsilon_t) = 0$ and $Avg(\omega_t) = 0$;
3) the predicted values for $\sigma_t, \lambda_t$ and cost of capital $R_t$ do not change during the recursive process;

According to conditions (1) and (2) the predicted error estimate $E_{t|t-1}$ is a white noise process with zero average. Indeed, condition (1) implies $Avg(V_{t|t-1}) = Avg(V_t)$ hence $Avg(E_{t|t-1}) = 0$. Furthermore, since by Eqs. (3.6) and (3.8) $E_{t|t-1}$ is a sum of uncorrelated white noise processes $\varepsilon_1, \varepsilon_2, \ldots, \varepsilon_t, \omega_1, \omega_2, \ldots, \omega_{t-1}$ hence the sum is also a white noise process with zero average. Consequently, the measurement residual $Res_t$ given by

$$Res_t = W_t - h.V_{t|t-1} = h.E_{t|t-1} + \lambda_t.\omega_t \qquad (6.1)$$

is also white noise with zero average because by Eq. (6.1) it is the sum of uncorrelated white noise processes $\varepsilon_1, \varepsilon_2, \ldots, \varepsilon_t, \omega_1, \omega_2, \ldots, \omega_t$. Since by condition (2) the measurement noise $\omega_t$ is uncorrelated $Covar(E_{t|t-1}, \omega_t) = 0$, the variance of $Res_t$ is

$$Var(Res_t) = h^2.Var(E_{t|t-1}) + {\lambda_t}^2 = {\lambda_t}^2.(1 + Q_t) \qquad (6.2)$$

where Eq. (3.17) has been used to express $Var(Res_t)$ in terms of $Q_t$ratio,. It is essential to note that condition (1) requires that the predicted cost of capital equals the actual cost (but unknown) of capital. To show this, let ${R_t}^*$ be the predicted cost of capital and let $R_t$ be the actual cost of capital. Given the updated company value $V_{t-1|t-1}$ at time t-1, then the predicted company value $V_{t|t-1}$ at time t is given by

$$V_{t|t-1} = (1 + {R_t}^*).V_{t-1|t-1} - FCFF_{t-1} \qquad (6.3)$$

Because $FCFF_t$ is non-stochastic, subtracting Eq. (6.3) for the predicted value $V_{t|t-1}$ from the actual value $V_t$ given by Eq. (2.1) the predicted error estimate $E_{t|t-1} = V_t - V_{t|t-1}$ obeys the relation

$$E_{t|t-1} = (1 + R_t).V_{t-1} - (1 + {R_t}^*).V_{t-1|t-1} + \sigma_t.\varepsilon_t \qquad (6.4)$$

which can be rewritten as

$$E_{t|t-1} = (R_t - {R_t}^*).V_{t-1} + (1 + {R_t}^*).E_{t-1|t-1} + \sigma_t.\varepsilon_t \qquad (6.5)$$

After inserting Eq. (3.7) into Eq. (6.5) the following recursion relation for $E_{t|t-1}$ is obtained

$$E_{t|t-1} = (R_t - R_t^{*}).V_{t-1} + (1 + R_t^{*}).(1 - h.k_{t-1}).E_{t-1|t-2} -$$
$$(1 + R_t^{*}).k_{t-1}.\lambda_{t-1}.\omega_{t-1} + \sigma_t.\varepsilon_t \qquad (6.6)$$

Therefore, as long as the predicted cost of capital $R_t^{*} = R_t$ and condition (3) is satisfied during the whole recursion process, then the first term on the RHS of Eq. (6.6) will vanish and by Eq. (6.1) both $E_{t|t-1}$ and $Res_t$ are white noise processes.

However, in case of changing market conditions such that $R_t \neq R_t^{*}$ the error estimate $E_{t|t-1}$ is not white noise due to the additional term $(R_t - R_t^{*}).V_{t-1}$ on the RHS of Eq. (6.6). This additional term may lead to nonzero average $Avg(E_{t|t-1})$ and therefore to divergence of the predicted value $V_{t|t-1}$ from the actual value $V_t$, thus violating condition (1) and increasing valuation risk $Var(E_{t|t-1})$. This divergence leads to systematic errors in the conventional Kalman filter and to reduced performance of the two-step filtering algorithm. Moreover, the increase of valuation risk $Var(E_{t|t-1})$ causes the Kalman gain $k_t \to 1$ and the Kalman filter to rely more and more on the measurements rather than the prediction model.

## 7. Elimination of systematic errors by adaptive Kalman filtering

In this section an adaptive Kalman filtering algorithm is proposed which not only detects systematic errors but also adjusts the predicted cost of capital $R_t^{*}$ and eliminates systematic errors such that the adjusted value $R_t^{*}$ converges to the actual value $R_t$.

Let $T_R$ be the typical time scale on which the actual cost of capital $R_t$ is varying, then on this time scale the cost of capital $R_t$ may be as approximately constant, say $R_t = R$. Next introduce the adjusted cost of capital $R_{Adj,t}$ such that the predicted cost of capital $R_t^{*}$ at time t equals the adjusted cost of capital $R_{Adj,t-1}$, at time t-1,

$$R_t^{*} = R_{Adj,t-1} \qquad (7.1)$$

The adjusted cost of capital $R_{Adj,t}$ at time t is defined as the weighted average of the $R_{Adj,t-1}$ and $R$,

$$R_{Adj,t} = w.R + (1 - w).R_{Adj,t-1} = R_{Adj,t-1} + w.(R - R_{Adj,t-1}) \qquad (7.2)$$

where $w$ is the so-called adjustment parameter such that $0 < w \leq 1$. Note that Eq. (7.2) implies that the difference $\Delta_t \equiv R - R_{Adj,t}$ between the adjusted cost of capital $R_{Adj,t}$ and the actual cost of capital $R$ obeys the recursion relation

$$\Delta_t = (1 - w).\Delta_{t-1} \qquad (7.3)$$

which by assumption $0 < w \leq 1$ implies that $\lim_{t\to\infty} \Delta_t = 0$ and consequently,

$$\lim_{t\to\infty} R_{Adj,t} = R \qquad (7.4)$$

Since convergence of $R_{Adj,t}$ to $R$ occurs on time scale $T_{Adj} = 1/w$ and $R_t$ is approximately constant on time scale $T_R$ this requires $T_{Adj} < T_R$ or equivalently $w > 1/T_R$. Therefore, the adjustment parameter w is arbitrary and small but must be larger than or $1/T_R < w \leq 1$. In case a relatively large value of w is selected then the adjustment can be relatively large and unstable, hence it is recommended to select small values of w within the above range.

The recursive formula Eq. (7.2) is not yet useful since it still depends on the unknown cost of capital $R$. This problem can be avoided by taking averages of LHS and RHS of Eq. (6.6) and using Eq. (7.1) to obtain

$$AVG(E_{t|t-1}) = (R - R_{Adj,t-1}).AVG(V_{t-1})$$
$$+(1 + R_{Adj,t-1}).(1 - h.k_{0,t-1})).AVG(E_{t-1|t-2}) \quad (7.5)$$

Here the moving average $\mathrm{AVG}(X_t)$ for the arbitrary time-dependent variable $X_t$ is defined as

$$AVG(X_t) \equiv \frac{1}{T}.\sum_{k=0}^{T-1} X_{t-k} \quad (7.6)$$

where $T$ is the averaging time scale. Assuming that the averaging time scale $T$ is much smaller than the typical time scale $T_R$ for changes in the cost of capital $R_t$, then $AVG(E_{t|t-1})$ is a slowly varying function of t and it is allowed to replace $AVG(E_{t|t-1})$ by $AVG(E_{t-1|t-2})$ on the LHS of Eq. (7.5). Solving the resulting expression for $R - R_{Adj,t-1}$ yields

$$R - R_{Adj,t-1} = \frac{[1 - (1 + R_{Adj,t-1}).(1 - h.k_{0,t-1})].AVG(E_{t-1|t-2})}{AVG(V_{t-1})} \quad (7.7)$$

Inserting Eq. (7.7) into Eq. (7.2) and upon replacing

$$AVG(Res_t) = h.AVG(E_{t|t-1}), \qquad AVG(W_t) = h.AVG(V_t) \quad (7.8)$$

it readily follows

$$R_{Adj,t} = R_{Adj,t-1} + w.\frac{[1 - (1 + R_{Adj,t-1}).(1 - h.k_{0,t-1})].AVG(Res_{t-1})}{AVG(W_{t-1})} \quad (7.9)$$

Note that the RHS of Eq. (7.9) is independent of $R$ and only depends on variables at time t-1 then Eq. (7.9) can be used to determine the adjusted cost of capital $R_{Adj,t}$ at time t. Since $R_{Adj,t}$ converges to $R$ on time scale $T_{Adj} = 1/w$ and consequently $AVG(Res_t)$ vanishes, the adaptive Kalman filter reduces to the conventional Kalman filter on the same time scale.

Since after recursively adjusting the cost of capital, systematic errors are eliminated and the measurement residue $Res_t$ is white noise, the two-step Kalman filtering algorithm can be applied as in section 3 in order to minimize valuation risk. An adaptive Kalman filtering algorithm is proposed by extending the conventional two-step Kalman algorithm by including a third step, the so-called parameter adjustment step which carried out before the value prediction step and the measurement update step. The three-step algorithm for the adaptive Kalman filter can be summarized as follows

**1. Parameter adjustment step**

Given the adjusted cost of capital $R_{Adj,t-1}$, Kalman gain $k_{0,t-1}$, measurement $W_{t-1}$ and measurement residual $Res_{t-1} = W_{t-1} - h.V_{t-1|t-2}$ at time t-1 the averages $AVG(W_{t-1})$ and $AVG(Res_{t-1})$ are determined. The adjusted cost of capital $R_{Adj,t}$ at time t can be recursively calculated by Eq. (7.8)

$$R_{Adj,t} = R_{Adj,t-1} + w.\frac{[1 - (1 + R_{Adj,t-1}).(1 - h.k_{0,t-1})].AVG(Res_{t-1})}{AVG(W_{t-1})} \quad (7.10)$$

**2. Value prediction step**

Given the updated company value $V_{t-1|t-1}$ and adjusted cost of capital $R_{Adj,t}$ given by step 1, the predicted company value $V_{t|t-1}$ at time t is determined as follows

$$V_{t|t-1} = (1 + R_{Adj,t}).V_{t-1|t-1} - FCFF_{t-1} \quad (7.11)$$

**3. Measurement update step**

From the measurement $W_t$ at time t and predicted value $V_{t|t-1}$ determined in step 2, the measurement residual can be determined at time t

$$Res_t = W_t - h.V_{t|t-1} \quad (7.12)$$

As in section 3 the updated company value $V_{t|t}$ is given by the weighted average of the predicted value $V_{t|t-1}$ and the measurement residual,

$$V_{t|t} = V_{t|t-1} + k_{0,t}.Res_t \quad (7.13)$$

The Kalman gain $k_{0,t}$ is determined as follows (cf. Eq. (3.15))

$$h.k_{0,t} = \frac{\mathrm{h}^2.VAR(Res_t)}{\mathrm{h}^2.VAR(Res_t) + \lambda^2} \quad (7.14)$$

where the variance is defined as

$$VAR(X_t) \equiv \frac{1}{T}.\sum_{k=0}^{T-1}(X_{t-k} - AVG(X_t))^2 \quad (7.15)$$

and $T$ is the averaging time scale. In Table 2 the main formulas are given for the FCFF model with the three-step adaptive Kalman filtering algorithm.

| |
|---|
| Step 1. Parameter adjustment step<br>(7.9) $R_{Adj,t} = R_{Adj,t-1} + w.\frac{[1-(1+R_{Adj,t-1}).(1-h.k_{0,t-1})].AVG(Res_{t-1})}{AVG(W_{t-1})}$ |
| Step 2. Value prediction step<br>(7.11) $V_{t\|t-1} = (1+R_{Adj,t}).V_{t-1\|t-1} - FCFF_t$ |
| Step 3. Measurement update step<br>(7.12) $Res_t = W_t - h.V_{t\|t-1}$<br>(7.13) $V_{t\|t} = V_{t\|t-1} + k_{0,t}.Res_t$<br>(7.14) $h.k_{0,t} = \frac{\mathrm{h}^2.VAR(Res_t)}{\mathrm{h}^2.VAR(Res_t)+\lambda^2}$ |

Table 2. Summary of the recursive formulas for the FCFF model with three-step adaptive Kalman filtering.

## 8. Monte Carlo simulation of the FCFF model with adaptive Kalman filtering

The performance of the adaptive Kalman filtering algorithm has been tested by a Monte Carlo simulation. In Table 3 the outcomes at different time steps t=10, t=50 and t=100 are given of a Monte Carlo simulation with 1.000 different realizations for a 2-period FCFF model with adaptive Kalman filtering and adjustment parameter $w = 0{,}05$. The FCFF model has forecast horizon $H = 40$ and free cash flows $F_I = 10{,}0$ in period I and $F_{II} = 7{,}0$ in period II. The predicted cost of capital at time t=0 is $R^* = 0{,}05$ and the actual (but unknown) cost of capital is $R = 0{,}1$. The noise statistics are $\sigma = 0{,}5$, $\lambda = 0{,}5$, $h = 1$. The averaging time scale for calculating of the average AVG and variance VAR is $T = 10$. The initial conditions at time t=0 are $V_{0|0} = W_0/h$ and $R_{Adj,0} = 0{,}05$.

| | Adaptive Kalman Filtering | | Conventional Kalman Filtering | |
|---|---|---|---|---|
| | Average | Stdev | Average | Stdev |
| $R_{adj}(10)$ | 7,23% | 0,15% | 5,00% | 0,00% |
| Res(10) | 2,9119 | 0,8640 | 5,4092 | 0,8402 |
| hk(10) | 0,9439 | 0,0205 | 0,9060 | 0,0276 |
| $R_{adj}(50)$ | 9,88% | 0,12% | 5,00% | 0,00% |
| Res(50) | 0,1329 | 0,8662 | 5,2554 | 1,2056 |
| hk(50) | 0,6825 | 0,1146 | 0,6858 | 0,1188 |
| $R_{adj}(100)$ | 10,00% | 0,12% | 5,00% | 0,00% |
| Res(100) | -0,0236 | 0,8723 | 5,4789 | 1,1680 |
| hk(100) | 0,6868 | 0,1114 | 0,6613 | 0,1120 |

Table 3. Statistics of a Monte Carlo simulation with 1000 different realizations for a 2-period FCFF model with adaptive Kalman filtering.

In Table 3 the statistics for the adaptive Kalman filtering are compared with the outcomes of the conventional Kalman filtering which clearly shows that the difference $R - R^*$ $l$eads to nonzero average residuals and systematic errors for all $t \geq 0$ in case of conventional Kalman filtering, whereas in case of adaptive Kalman filtering the average residuals vanish at time t=50 and t=100 which is much larger than the averaging time scale $T = 10$. Fig. 1 demonstrates that the adjusted cost of capital $R^*$ approximates the actual cost of capital $R = 0{,}1$ on adjustment timescale $\mathrm{T_{Adj}} = 1/\mathrm{w} = 20$ .

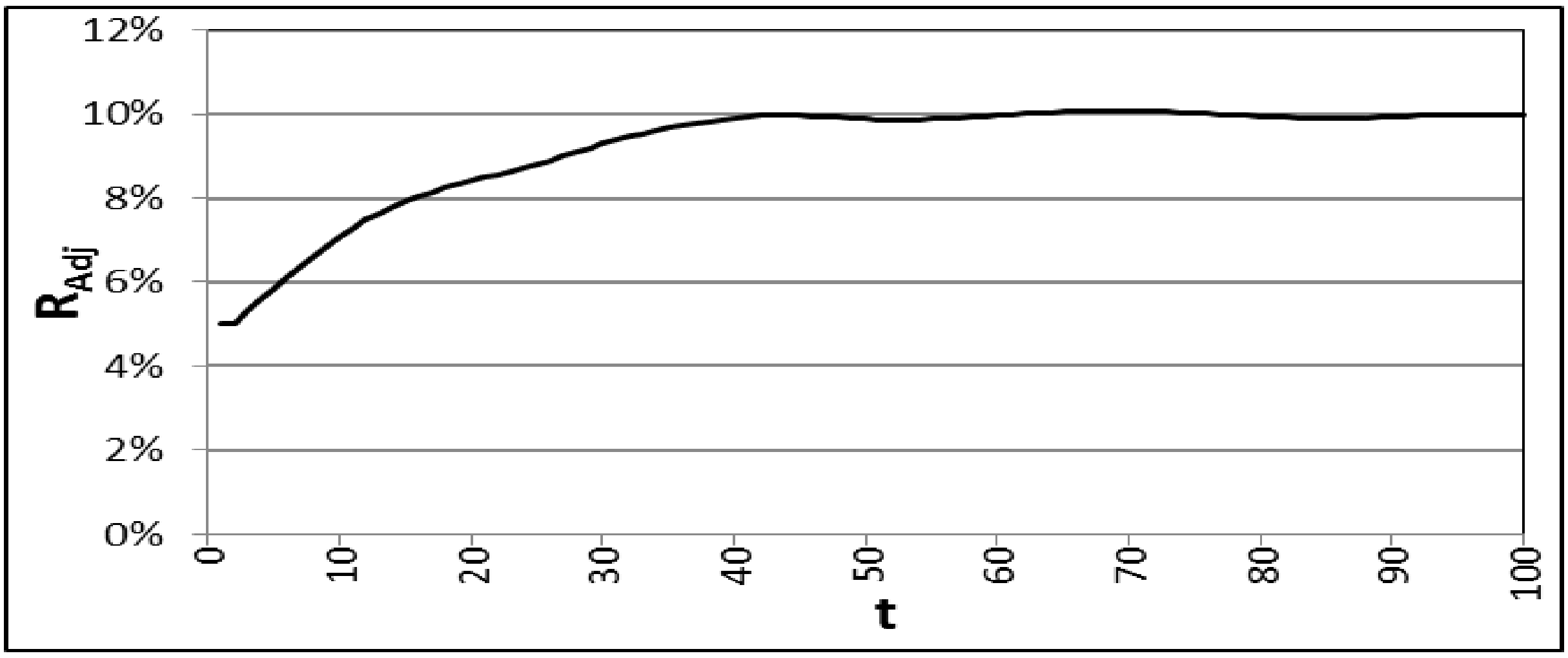

Fig 6. Adjusted cost of capital $R_{Adj,t}$ as a function of time

## 9. Conclusion

In this research paper a recursive free cash flow FCFF model is proposed to determine the corporate value in an efficient market in which new market and company-specific information is modelled by additive white noise. By implementing a conventional Kalman filter into the FCFF model, it has been pointed out that valuation risk can be reduced significantly, thus improving the predictive power of the FCFF model. Applying conventional Kalman filtering is adequate as long as the predicted parameters of the valuation model match the actual market parameters. It is shown that conventional Kalman filtering may exhibit divergence behavior and systematic errors in case of market parameters are changing and deviate from the model parameters. An adaptive Kalman filtering is proposed which not only eliminates the systematic errors but also adaptively adjusts the predicted value of the cost of capital such that it converges to the actual cost of capital. The performance of the adaptive Kalman filter has been tested by Monte Carlo simulation demonstrating the reliability and robustness of the algorithms to reduce valuation risk and adjusting the cost of capital. In this research paper the conventional and adaptive Kalman filter has been implemented into the FCFF model but in the Appendix it has been pointed out that the algorithm can also be implemented into other valuation models such as the economic value added (EVA™) model and free cash flow to equity (FCFE) models.

## Appendix A

In section 4 it was pointed out that convergence for $t \to \infty$ of the time series for $Var(E_{t|t})$ requires that $a = (1 - h.k).(1 + R) < 1$. At the optimum Kalman gain $k = k_0$ given by Eq. (3.18) the parameter $a$ can be written as

$$a = \frac{1 + R}{1 + Q} \tag{A.1}$$

where $Q$ is given by Eq. (5.4)

$$Q = \frac{\Pi + \sqrt{\Pi^2 + 4x^2}}{2x^2} \tag{A.2}$$

where $\Pi \equiv 1 + x^2.(2R + R^2)$ and $x = \lambda/h\sigma$ is the dimensionless parameter determined by the ratio of the measurement noise $\lambda$ and process noise $\sigma$. Eq. (A.1) implies that convergence $a < 1$ requires that $Q > R$ . It will now be proved that for arbitrary but fixed value of R, the parameter $Q > R$ for all $x \geq 0$. Inserting the expression for $\Pi \equiv 1 + x^2.(2R + R^2)$

$$Q = \frac{1 + x^2.(2R + R^2) + \sqrt{1 + 2x^2.(2 + 2R + R^2) + x^4.(2R + R^2)^2}}{2x^2} \tag{A.3}$$

Asymptotic values of $Q$ for $x \ll 1$ and for $x \gg 1$ are obtained by Taylor expansion of Eq. (A.3)

$$Q \approx \frac{1}{x^2} \; for \; x \ll 1$$

$$Q \approx 2R + R^2 \;\; for \; x \gg 1 \tag{A.4}$$

By taking the derivative of Eq. (A.3) with respect to $x$ for arbitrary but fixed $R$, the ratio $Q$ is a decreasing function for all $x \geq 0$

$$Q'(x) = -\frac{1}{x^3} - \frac{1 + x^2.(2 + 2R + R^2)}{x^3.\sqrt{1 + 2x^2.(2 + 2R + R^2) + x^4.(2R + R^2)^2}} < 0 \tag{A.5}$$

Since $Q$ is a decreasing function of $x$ for all $x > 0$ with asymptotic values $Q \approx 1/x^2 \gg R$ for $x \ll 1$ and $Q \approx 2R + R^2 > R$ for $x \gg 1$, this proves that $Q > R$ for all $x \geq 0$.

Asymptotic expressions for $h.k_0$, $Var(E_{\infty|\infty})$ and $\mathcal{R}$ are obtained by inserting Eq. (A.4) into Eqs. (5.5), (5.6) and (5.8)

For $x \ll 1$: $\quad h.k_0 \approx 1 - x^2 \quad Var(E_{\infty|\infty}) \approx \sigma^2.x^2 \quad \mathcal{R} \approx (R^2 + 2R).x^2$

(A.6)

For $x \gg 1$: $\quad h.k_0 \approx 2R \quad Var(E_{\infty|\infty}) \approx 2R.\sigma^2.x^2 \quad \mathcal{R} \approx 2R.(R^2 + 2R).x^2$

Note that for $x > 1$ Kalman filtering algorithm is effective as long as $\mathcal{R} < 1$ hence $x < 1/\sqrt{2R.(R^2 + 2R)}$ by Eq. (A.6). Therefore, the algorithm is effective even in the range $1 < x < 1/\sqrt{2R.(R^2 + 2R)}$ where the measurement noise is larger than the process noise. This range will be larger the smaller is the value of $R$ as illustrated by Fig. 5.

## Appendix B

In previous sections the Kalman filtering algorithm has been implemented into the FCFF model. It will now be shown that the Kalman filter algorithm can also be implemented in other univariate valuation models such as the economic value added (EVA™) model and free cash flow to equity (FCFE) model after converting these into recursive models.

In order to determine the recursion equation for the EVA™ model, a registered trademark developed by US consulting firm of Stern Stewart [3], start with the univariate valuation model given by Eq. (2.1)

$$V_t = (1 + R_t).V_{t-1} - FCFF_{t-1} + \sigma_t.\varepsilon_t \qquad \text{(B.1)}$$

Recalling that the free cash flow $FCFF_{t-1}$ equals the net operating profit after tax $NOPAT_{t-1}$ realized in the time period between time t-1 and time t minus the net investment in operating invested capital $\Delta OIC_t = OIC_t - OIC_{t-1}$

$$FCFF_{t-1} = NOPAT_{t-1} - OIC_t + OIC_{t-1} \qquad \text{(B.2)}$$

which upon substituting into Eq. (B.1) gives

$$V_t = (1 + R_t).V_{t-1} - NOPAT_{t-1} + OIC_t - OIC_{t-1} + \sigma_t.\varepsilon_t \qquad \text{(B.3)}$$

Next define the Shareholder Value Added at time t, denoted by $SVA_t$

$$SVA_t = V_t - OIC_t \qquad \text{(B.4)}$$

then it readily follows from Eq. (B.3) that

$$SVA_t = (1 + R_t).SVA_{t-1} - EVA_{t-1} + \sigma_t.\varepsilon_t \qquad \text{(B.5)}$$

where Economic Value Added $EVA_t$ is defined as

$$EVA_{t-1} = NOPAT_{t-1} - R_t.OIC_{t-1} = (ROIC_t - R_t).OIC_{t-1} \qquad \text{(B.6)}$$

Therefore $EVA_t$ equals the net operating profit after tax less the cost of operating capital invested in the business. In other words, $EVA_t$ expresses the fundamental principle that the company is creating (destroying) shareholder value when the return on operating invested capital $ROIC_t = NOPAT_t/ OIC_t$ is larger (smaller) than the weighted average cost of capital $R_t$, which can be interpreted as the hurdle rate for shareholder value creation. Since Eq. (B.5) has the same functional form as Eq. (2.1) implementation of the Kalman filtering algorithm is straightforward by replacing $FCFF_t \rightarrow\ EVA_t$ and $V_t \rightarrow\ SVA_t$ in previous sections.

The Kalman filtering algorithm can also be implemented into the FCFE model which is a more direct way to determine the shareholder value or equity value $\mathrm{E_t}$. Let $\mathrm{R_{e,t}}$ be the return on equity at time t and $FCFE_{t-1}$ be the free cash flow to equity realized in time period between time t-1, then $E_t$ obeys the following recursion formula

$$E_t = \left(1 + R_{e,t}\right).E_{t-1} - FCFE_{t-1} + \sigma_{e,t}.\varepsilon_t \qquad \text{(B.7)}$$

Implementation of the Kalman filtering algorithm then is straightforward by replacing $FCFF_t \rightarrow\ FCFE_t$ and $V_t \rightarrow\ E_t$.